\documentclass[conference]{IEEEtran}
\IEEEoverridecommandlockouts

\usepackage{cite}
\usepackage{amsmath,amssymb,amsfonts}
\usepackage{algorithmic}
\usepackage{graphicx}
\usepackage{textcomp}
\usepackage{xcolor}
\usepackage{graphicx}
\usepackage{booktabs}
\usepackage{tcolorbox}
\usepackage{hyperref}
\usepackage{multirow}

\def\BibTeX{{\rm B\kern-.05em{\sc i\kern-.025em b}\kern-.08em
		T\kern-.1667em\lower.7ex\hbox{E}\kern-.125emX}}
\begin{document}
	\title{Removing Noise or Introducing Bias? The Hidden Cost of MSR Filtering}
	
	\author{\IEEEauthorblockN{Mohit Kaushik}
		\IEEEauthorblockA{\textit{Department of Computer Science} \\
			\textit{Guru Nanak Dev University}\\
			Amritsar, India \\
			mohitcs.rsh@gndu.ac.in}
		\and
		\IEEEauthorblockN{Jyoti Bawa}
		\IEEEauthorblockA{\textit{Department of Computer Science} \\
			\textit{Guru Nanak Dev University}\\
			Amritsar, India \\
			jyoticsc.rsh@gndu.ac.in}
		\thanks{*Presented at the International Conference on Computing Innovations for Sustainability (ICCIS), Guru Nanak Dev University, Amritsar, India.}
	}
	
	\maketitle

\begin{abstract}

OSS platforms like GitHub serve as a primary data source for MSR research. As the platform is widely used by different users, spanning from student to developer, not all repositories are actual engineered projects. Therefore, to avoid such noise, researchers often apply several criteria, which may fundamentally change the sample demographics. To understand such biases, this study aims to uncover the hidden cost originating from these arbitrary thresholds or criteria. We analyzed 1.57 million repositories from the SEART platform and constructed several datasets from the thresholds often applied in MSR research. We identify the maintenance bias in these filtering processes, which masks the true abandonment (73.42\%) realities of OSS projects. Also, these strategies favor some ecosystems and governance styles. Moreover, the sampling strategy also distorts the relationship between variables, suffering from relational biases. Therefore, to avoid such biases, the researchers should shift towards stratified sampling and refining the criteria for noise detection.

\end{abstract}


\section{Introduction}

Open Source Software (OSS) platforms like GitHub are fueling the software engineering (SE) research. From an analysis of $\approx$10,000 SE papers, 20-25\% of papers actually utilize GitHub artifacts such as commits, PRs, Issues, stars, etc~\cite{alrashedy2024software}. This proportion is even larger for top mining software repository (MSR) venues~\cite{alrashedy2024software}. Approximately 40-50\% of studies from venues like ICSE, ASE, and FSE use GitHub data~\cite{alrashedy2024software}. As of 2026, GitHub officially hosts 630 million repositories~\cite{github_octoverse2025}, which make it the primary data source for SE or MSR studies. But not all repositories are a valid data source for research.

GitHub is widely used by students, teachers, or instructors for experimentation purposes~\cite{kalliamvakou2016depth,munaiah2017curating}. As a result, the researchers usually exclude these experimental or toy repositories from their analysis~\cite{dabic2021sampling}. Considering such repositories as noise, the researchers usually apply certain noise-removing filters~\cite{kalliamvakou2016depth}. These filters include, but are not limited to, popularity-based or activity-based filters~\cite{munaiah2017curating}. For example, some studies use a generic threshold of a minimum of 10 stars (a proxy for project popularity on GitHub) to remove noise~\cite{dabic2021sampling}. But this 10-star baseline still leaves massive noise. To further handle this noise, the researchers apply a secondary, higher star threshold, for example, selecting the repositories having more than 500 stars~\cite{munaiah2017curating}. Although it actually helps to combat noise, it completely changes the sample demographics, for instance, favoring a certain governance style or ecosystem~\cite{malviya2024role}. Moreover, it can change the statistical relationship, thereby limiting the results' generalizability to their own platform~\cite{baltes2022sampling}.

Therefore, this study aims to analyze the change in sample demographics by these arbitrary filters. In this study we answer how these noise-reduction thresholds do not just clean the data but systematically distort the reality of the OSS ecosystem. Specifically, we aim to analyze the composition, maintenance, and relational biases that originated from these arbitrary thresholds.

\section{Related Works}

Historically, SourceForge was the primary data source for MSR research. Foundational studies such as those by Howison and Crowston~\cite{howison2004perils} highlighted some metadata inconsistencies in SourceForge projects. The SourceForge projects suffered from survivorship bias, where only 1\% of projects were actually active~\cite{rainer2005evaluating}. Later on, GitHub was introduced in 2008. The prime GitHub data sources, such as GHTorrent~\cite{gousios2012ghtorrent} and GHArchive~\cite{grigorik2012gharchive}, were introduced and launched in 2012. These platforms host thousands to millions of repositories' data, thereby providing ease to the MSR researchers. However, some researchers were blindly treating all GitHub repositories as legitimate software engineering projects. 

Kalliamvakou et al.'s study~\cite{kalliamvakou2016depth}, presented in MSR 2014, noted 13 specific perils (dangers) to blindly mining or using GitHub for research. They specifically highlighted that 37\% of analyzed projects were actually noise or experimental and a vast majority of projects were dead. Following their established perils, the MSR researchers are now applying specific noise filters, such as setting a threshold on project stars, commits, pull requests, etc~\cite{munaiah2017curating,dabic2021sampling}. However, these filters introduce severe biases. Tutko et al.~\cite{tutko2022software}, in their systematic literature review of 286 recent MSR studies, highlighted that $\approx$86.7\% of MSR papers rely on non-probability sampling. 43.7\% of the sampled papers utilized the popularity-based criteria to avoid noise or trivial repositories.

Despite the high popularity of star-based filters, they suffer from severe data loss. To demonstrate this, Munaiah et al.~\cite{munaiah2017curating} conducted a study on identifying real software engineer projects. They highlighted that using a 10-star threshold exhibited a high precision of 97\% with a terrible recall of only 32\%. The authors further tested their model on 50 and 500 star thresholds and noted the 14\% and 0\% recall. Therefore, relying on a star-based threshold to identify a real software engineering project is risky, as the same authors highlighted a loss of 68\% of genuine software projects. Even our pilot study using GitHub's advanced search found 317 million repositories that have fewer than 10 stars. Our manual check found that some of them were highly active with more than 500 commits. Additionally, the number of the stars on a project does not equate to their actual usage. Koch et al.~\cite{koch2024fault} revealed a weak correlation between number of stars and actual downloads. As a result, the star-based thresholds are lossy and misleading in nature.

Considering the potential biases with star-based thresholds, some studies adopted activity-based criteria. For instance, Xiao et al.~\cite{xiao2023early} applied a 95th-percentile activity threshold by excluding the repositories having fewer than 57 commits. Similarly, Ray et al.~\cite{ray2014large} also retain the repositories satisfying the threshold of 20 commits to isolate the active projects from noise. Raw commit volume suffers from accumulation bias and is further inflated by bots~\cite{dey2020exploratory}. Kaushik and Chahal's study~\cite{kaushik2026} specifically identified that monthly rates of commits are highest in younger, short-lived projects and relying on higher commit-activity thresholds, the researchers are sampling short-lived projects that have an initial burst of commit activity. Therefore, the MSR studies should not rely on the assumption that high commit activity is always associated with longer-lived and healthier projects~\cite{kaushik2026}. Moreover, these commit-based thresholds not only clean the dataset by removing experimental noise but also change the statistical relationship between variables. Thakur and Mockus~\cite{malviya2024role} revealed a flip in the relationship between project activity and popularity after applying stricter commit-based thresholds. As these thresholds inadvertently change the relationship between variables, models derived from these thresholds may introduce bias or limit generalizability.

\section{Methodology}
This section presents our research questions, data source, and methodology.
\subsection{Research Questions}
Our study is guided by three research questions, outlined below.
\begin{itemize}
	\item RQ1: How do popularity and activity-based thresholds alter the representation of programming language, repository age, and project size?
	\item RQ2: How do the arbitrary commit and star thresholds change the active project age and project abandonment rate?
	\item RQ3: To what extent do these noise filters change the statistical relationship between developer activity and project popularity?
	
\end{itemize}

\subsection{Data Collection, Preprocessing and Construction}

To initiate the analysis, we collected the dataset from the SEART platform\footnote{https://seart-ghs.si.usi.ch}. This platform was introduced by Dabic et al.~\cite{dabic2021sampling} in MSR 2021. To handle noise and support valid repository sampling, the platform natively applied a 10-star threshold. We downloaded a dataset of 1.9 million repositories across 35 attributes. Following this, we removed the null values for commits, total issues, and contributors. Moreover, we removed the forked repositories to avoid duplication and excluded empty projects (size = 0 KB). This process yielded a clean baseline dataset of 1.57 million repositories.

\begin{figure}[!h]
	\centering
	\includegraphics[width=\linewidth]{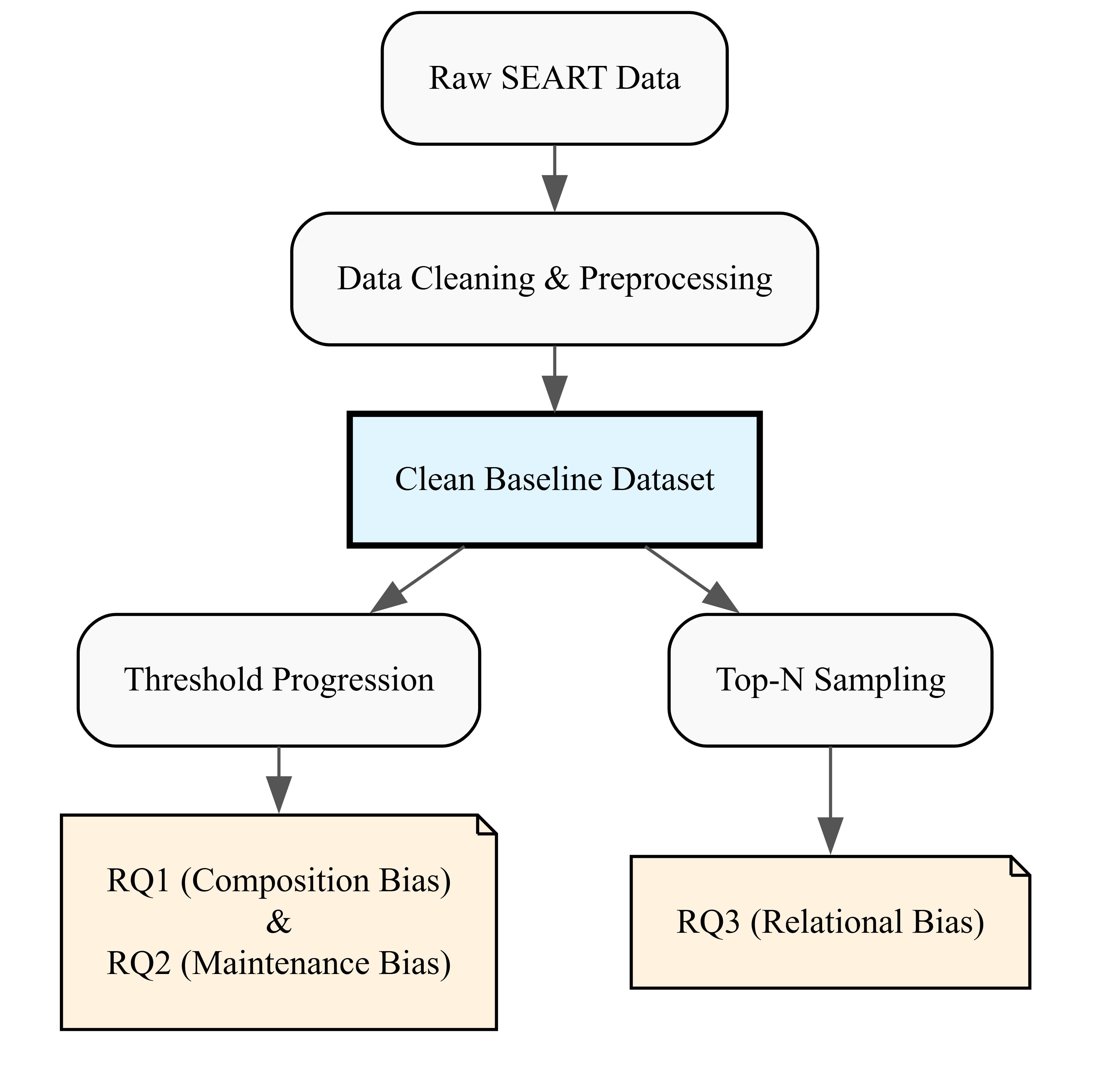}
	\caption{An Overview of the research approach followed in this work}
	\label{fig:methodology}
\end{figure}

 For dataset construction, we applied both popularity- and activity-based thresholds. As illustrated in Figure~\ref{fig:methodology}, we followed a two-pronged approach. For RQ1 and RQ2, we applied various thresholds on the baseline dataset, outlined below.

\begin{itemize}
	\item Popularity-based thresholds: We tested the sample composition and maintenance bias by varying the star-based thresholds from 10 to 1000 (10, 20, 50, 100, 500, 1000). With these thresholds, we computed the projects' programming language compositions, license proportions, median size, median team size, and median age.
	\item Activity-based thresholds: Similarly, we varied the commit threshold from 10 to 1000 (10, 20, 50, 100, 500, 1000).
\end{itemize}

For RQ3, we constructed the three datasets: Dataset A, B, and C. Dataset A contains the top 10,000 most popular projects, whereas Dataset B contains the top 10,000 most active projects (top commits). Dataset C also contains 10,000 randomly sampled repositories stratified by programming language . All these datasets were constructed from a baseline dataset of 1.57 million repositories. To answer RQ3, we compared these datasets’ attributes' correlations to reveal how these criteria inflate or degrade the correlation between the underlying attributes.

\section{Results}
This section presents the results of this study. We organized these results in accordance with the proposed research questions. Table~\ref{tab:threshold_progression_combined} summarizes the results for RQ1 and RQ2.

\begin{table*}
	\centering
	\caption{Threshold progression analysis for popularity (stars) and activity (commits)}
	\label{tab:threshold_progression_combined}
	
	\begin{tabular}{llcccccr}
		\toprule
		\textbf{RQ} & \textbf{Metric} & \multicolumn{6}{c}{\textbf{Stars Thresholds}} \\
		\cmidrule(lr){3-8}
		&  & $\geq 10$ & $\geq 20$ & $\geq 50$ & $\geq 100$ & $\geq 500$ & $\geq 1000$ \\
		\midrule
		\multirow{3}{*}{RQ1} 
		& Size (KB)       & 1,360.0 & 1,719.0 & 2,420.0 & 3,221.0 & 6,648.5 & 9,477.0 \\
		& Contributors    & 2       & 3       & 4       & 6       & 14      & 21      \\
		& Age (Months)    & 26.74   & 32.29   & 39.26   & 44.94   & 60.25   & 67.05   \\
		\midrule
		\multirow{5}{*}{RQ2} 
		& Permissive (\%) & 50.12   & 52.91   & 56.01   & 58.15   & 62.24   & 63.18   \\
		& No License (\%) & 30.03   & 26.19   & 21.85   & 18.75   & 12.20   & 10.18   \\
		& Copyleft (\%)   & 11.01   & 11.26   & 11.58   & 11.89   & 12.59   & 12.93   \\
		& Inactive (\%)   & 73.42   & 71.13   & 67.79   & 64.73   & 55.52   & 50.65   \\
		& Creation Date   & 2020-02 & 2019-09 & 2019-04 & 2019-01 & 2018-05 & 2018-02 \\
		\bottomrule
	\end{tabular}
	
	\vspace{0.5em} 
	
	\begin{tabular}{llcccccr}
		\toprule
		\textbf{RQ} & \textbf{Metric} & \multicolumn{6}{c}{\textbf{Commits Thresholds}} \\
		\cmidrule(lr){3-8}
		&  & $\geq 10$ & $\geq 20$ & $\geq 50$ & $\geq 100$ & $\geq 500$ & $\geq 1000$ \\
		\midrule
		\multirow{3}{*}{RQ1} 
		& Size (KB)       & 1,548.0 & 1,876.0 & 2,846.0 & 4,387.0 & 15,517.0 & 28,335.5 \\
		& Contributors    & 3       & 3       & 4       & 5       & 12       & 19       \\
		& Age (Months)    & 29.50   & 32.88   & 39.19   & 45.20   & 61.32    & 68.63    \\
		\midrule
		\multirow{5}{*}{RQ2} 
		& Permissive (\%) & 51.49   & 52.65   & 53.85   & 53.96   & 50.61    & 47.63    \\
		& No License (\%) & 27.92   & 25.72   & 22.17   & 19.63   & 15.32    & 13.91    \\
		& Copyleft (\%)   & 11.47   & 12.08   & 13.39   & 14.70   & 18.41    & 20.22    \\
		& Inactive (\%)   & 72.12   & 69.97   & 65.03   & 59.82   & 45.80    & 40.49    \\
		& Creation Date   & 2020-01 & 2019-12 & 2019-11 & 2019-10 & 2019-04  & 2018-11 \\
		\bottomrule
	\end{tabular}
\end{table*}

Popularity-based thresholds (Table~\ref{tab:threshold_progression_combined}, upper portion) highlight the governance and maintenance biases. As the popularity-based threshold increases, the project size increases substantially by 7x. This indicates that a sample of  higher-popularity projects favors large projects. In a similar manner, median contributors’ size and project age also scale gradually. Median creation year also revealed a shift from 2020 to 2018. Regarding maintenance bias, the baseline noted 73.42\% inactive projects favoring the active projects by continuously reducing the inactive projects' proportions. Moreover, the higher popularity thresholds favor permissive projects, continuously dropping the informal, ad hoc projects with no license. In the case of ecosystem share, Python (~20\%) and JavaScript (13-14\%) hold steadily as the threshold progresses. TypeScript's proportion increased from 9 to 12\%. Go, which is not in the top five programming languages at lower thresholds (10-20 stars), increased its proportion to 7.6\% at 1000 stars. Conversely, the C++ present in lower thresholds ($\leq$ 50)  drops out of the top 5 after $\geq$ 500 stars. Overall, this analysis confirms that the higher popularity thresholds favor older, large, permissive, and infrastructural projects, which attract higher popularity and longevity.

Similarly, for median project size, team size, and project age, the commit-based threshold (Table~\ref{tab:threshold_progression_combined}, lower portion) revealed a similar behavior. As the commit threshold increases, the project size is increased by $\approx$18x. Moreover, the median number of contributors and project age also steadily increased from 3 to 19 and 29.50 months to 68.63 months, respectively. Regarding governance, the commit-based threshold favors the copyleft license by dropping permissive license shares from 51.49\% to 47.63\%. Copyleft’s share is increased from 11.47\% to 20.22\% on higher thresholds. Moreover, inactive projects’ proportions continuously dipped from 72.12\% to 40.49\%. These activity-based thresholds also favor certain ecosystems. Python (20\%-17\%) and JavaScript (14\%-10.8\%) proportions gradually decline at higher thresholds. TypeScript (9.4\%-12.7\%) and Java (5.4\%-8.4\%) proportions increased steadily with higher commit thresholds. C++ nearly doubled its share from 5.36\% to 9.51\%. Similar to star-based thresholds, the commit-based thresholds also favor older, large infrastructural projects with larger teams but support restrictive copyleft licenses. Table~\ref{tab:correlation_results} reports the Spearman correlations for the developer activity and popularity attributes across all three datasets. We applied Spearman because the OSS dataset, specifically from GitHub, inherently exhibits a non-linear relationship, and the distribution-agnostic nature of Spearman makes it suitable for this dataset.

\begin{table*}
	\centering
	\caption{Spearman correlations across datasets (Relational Bias)}
	\label{tab:correlation_results}
	\begin{tabular}{lccr}
		\toprule
		\textbf{Variable Pair} & \textbf{Dataset C} & \textbf{Dataset A} & \textbf{Dataset B} \\
		& Baseline & Popularity Filter & Activity Filter \\
		\midrule
		commits vs. size             & 0.466 & 0.603↑ & 0.808↑ \\
		commits vs. totalIssues      & 0.517 & 0.707↑ & 0.331↓ \\
		commits vs. contributors     & 0.588 & 0.789↑ & 0.720↑ \\
		totalIssues vs. contributors & 0.568 & 0.692↑ & 0.365↓ \\
		totalIssues vs. stargazers   & 0.499 & 0.323↓ & 0.369↓ \\
		totalIssues vs. size         & 0.205 & 0.435↑ & 0.284↑ \\
		contributors vs. size        & 0.177 & 0.375↑ & 0.624↑ \\
		contributors vs. stargazers  & 0.359 & 0.321↓ & 0.228↓ \\
		commits vs. stargazers       & 0.282 & 0.234↓ & 0.243↓ \\
		size vs. stargazers          & 0.159 & 0.192↑ & 0.241↑ \\
		\bottomrule
	\end{tabular}
	
	\vspace{0.5em}
	\footnotesize{↑ = inflated correlation, ↓ = degraded correlation.}
\end{table*}

In the baseline (Dataset C), all the correlations are weak to moderate. However, filtering significantly warps these relationships. The correlation of commits with size is inflated in both popular and active projects. This is expected, as both thresholds favor large projects. Moreover, the relationships between commits and total issues as well as total issues and contributors are inflated in the popularity-filtered sample (0.707 and 0.692) but degraded in the activity-filtered sample (0.331 and 0.365). Also, the correlations with stargazers degrade independently of the filtering process. Overall, these findings confirmed the presence of relational bias. Applying top-N filtering fundamentally distorts the relationship between software metrics.

\section{Discussion and Limitations}

Our study aims to highlight the potential trade-offs between noise and bias. MSR studies usually adopt arbitrary thresholds to eliminate noise. These criteria, while successfully removing the noises, introduce several biases, which in turn may lead to a generalizability crisis~\cite{baltes2022sampling}. Our results highlight how adopting a filtering strategy favors certain ecosystems and governance, which masks the OSS ecosystem's messy realities. The statistical relationship varied substantially across these filtering techniques. If a study with top-N filtering trains a predictive model, the model will likely learn the sampling artifacts rather than actual software engineering principles and relationships. These findings are aligned with Kalliamvakou et al.~\cite{kalliamvakou2016depth}, who first highlight the OSS projects' reality by highlighting promise and perils. Similar concerns were also raised by Tutko et al.~\cite{tutko2022software} that removing noise fundamentally changes the relationship between variables.

\paragraph*{Limitations:}

Our study has certain limitations. First, we relied on the SEART platform for a data source, which may contain data biases. The dataset, which we considered as a baseline, natively satisfies the 10-star threshold, which completely avoids the absolute bottom tail of GitHub projects. Moreover, as our pilot study noted, unpopular repositories are not always noise. Second, we considered the commits and stargazers as proxies for activity and popularity. These metrics natively suffer from accumulation bias and are mediated by bots' presence~\cite{dey2020exploratory}.

\section{Conclusion and Future Work}

This study highlighted several severe biases originating from filtering processes. Our results highlighted that different thresholds distort the dataset demographics. Star-based thresholds artificially inflated the project size by 7x, whereas commit-based thresholds increased it by 18x. Additionally, star-based thresholds favored the permissive licenses and infrastructural projects. On the other hand, commit-based thresholds favored different governance styles. Moreover, both threshold criteria inflated or degraded the correlation, thereby introducing correlation bias. Due to these biases, the MSR researchers must stop applying these arbitrary, top-N thresholds without sensitivity analysis or proper justification. Instead, future research should adopt stratified random sampling approaches to ensure that removing noise does not destroy the sample representativeness. Moreover, the definition and criteria to identify noise could be refined to avoid dropping real engineered projects.

%
%
%
\bibliographystyle{IEEEtran}
\bibliography{references}

@INPROCEEDINGS{alrashedy2024software,
	author    = {Alrashedy, Kamel and Binjahlan, Ahmed},
	title     = {How do Software Engineering Researchers Use GitHub? An Empirical Study of Artifacts \& Impact},
	booktitle = {2024 IEEE International Conference on Source Code Analysis and Manipulation (SCAM)},
	year      = {2024},
	pages     = {118--130},
	publisher = {IEEE},
	doi       = {10.1109/SCAM63643.2024.00021},
	url={https://doi.org/10.1109/SCAM63643.2024.00021}
}

@misc{github_octoverse2025,
	author       = {GitHub Staff},
	title        = {Octoverse 2025: A new developer joins GitHub every second as AI leads TypeScript to \#1},
	year         = {2025},
	month        = {October},
	howpublished = {\url{https://gh.io/octoverse2025}},
	note         = {Accessed: May 23, 2026}
}

@article{kalliamvakou2016depth,
	author    = {Kalliamvakou, Eirini and Gousios, Georgios and Blincoe, Kelly and Singer, Leif and German, Daniel M. and Damian, Daniela},
	title     = {An in-depth study of the promises and perils of mining GitHub},
	journal   = {Empirical Software Engineering},
	volume    = {21},
	number    = {5},
	pages     = {2035--2071},
	year      = {2016},
	publisher = {Springer},
	doi       = {10.1007/s10664-015-9393-5}
}

@article{munaiah2017curating,
	author    = {Munaiah, Nuthan and Kroh, Steven and Cabrey, Craig and Nagappan, Meiyappan},
	title     = {Curating GitHub for engineered software projects},
	journal   = {Empirical Software Engineering},
	volume    = {22},
	number    = {6},
	pages     = {3219--3253},
	year      = {2017},
	publisher = {Springer},
	doi       = {10.1007/s10664-017-9512-6},
	url={https://link.springer.com/article/10.1007/s10664-017-9512-6}
}

@INPROCEEDINGS{dabic2021sampling,
	author    = {Dabic, Ozren and Aghajani, Emad and Bavota, Gabriele},
	title     = {Sampling Projects in GitHub for MSR Studies},
	booktitle = {2021 IEEE/ACM 18th International Conference on Mining Software Repositories (MSR)},
	year      = {2021},
	pages     = {560--564},
	publisher = {IEEE},
	doi       = {10.1109/MSR52588.2021.00074}
}

@inproceedings{malviya2024role,
	author    = {Malviya-Thakur, Addi and Mockus, Audris},
	title     = {The Role of Data Filtering in Open Source Software Ranking and Selection},
	booktitle = {Proceedings of the 1st IEEE/ACM International Workshop on Methodological Issues with Empirical Studies in Software Engineering},
	year      = {2024},
	pages     = {7--12},
	publisher = {Association for Computing Machinery},
	address   = {New York, NY, USA},
	doi       = {10.1145/3643664.3648210},
	url       = {https://doi.org/10.1145/3643664.3648210}
}

@article{baltes2022sampling,
	author    = {Baltes, Sebastian and Ralph, Paul},
	title     = {Sampling in software engineering research: A critical review and guidelines},
	journal   = {Empirical Software Engineering},
	volume    = {27},
	number    = {4},
	pages     = {94},
	year      = {2022},
	publisher = {Springer},
	doi       = {10.1007/s10664-021-10072-8},
	url={https://link.springer.com/article/10.1007/s10664-021-10072-8}
}

@inproceedings{howison2004perils,
	author    = {Howison, James and Crowston, Kevin},
	title     = {The perils and pitfalls of mining SourceForge},
	booktitle = {Proceedings of the 26th International Conference on Software Engineering - W17S Workshop: International Workshop on Mining Software Repositories (MSR 2004)},
	pages     = {7--11},
	year      = {2004},
	publisher = {IET},
	doi       = {10.1049/ic:20040467},
	url = {https://doi.org/10.1049/ic:20040467}
}

@inproceedings{rainer2005evaluating,
	author    = {Rainer, Austen and Gale, Stephen},
	title     = {Evaluating the Quality and Quantity of Data on Open Source Software Projects},
	booktitle = {Proceedings of the 1st International Conference on Open Source Systems (OSS 2005)},
	pages     = {29--36},
	year      = {2005},
	url       = {https://uhra.herts.ac.uk/id/eprint/13709/}
}

@inproceedings{gousios2012ghtorrent,
	author    = {Gousios, Georgios and Spinellis, Diomidis},
	title     = {GHTorrent: GitHub's Data from a Firehose},
	booktitle = {2012 9th IEEE Working Conference on Mining Software Repositories (MSR)},
	pages     = {12--21},
	year      = {2012},
	publisher = {IEEE},
	doi       = {10.1109/MSR.2012.6224294}
}

@misc{grigorik2012gharchive,
	author = {Ilya Grigorik},
	title = {GH Archive: GitHub Public Timeline Data},
	year = {2012},
	howpublished = {\url{https://www.gharchive.org/}},
	note = {Accessed: May 23, 2026}
}

@article{tutko2022software,
	author       = {Tutko, Adam and Henley, Austin Z. and Mockus, Audris},
	title        = {How are Software Repositories Mined? A Systematic Literature Review of Workflows, Methodologies, Reproducibility, and Tools},
	journal      = {arXiv preprint arXiv:2204.08108},
	year         = {2022},
	eprint       = {2204.08108},
	archivePrefix= {arXiv},
	primaryClass = {cs.SE},
	url          = {https://arxiv.org/abs/2204.08108}
}

@inproceedings{koch2024fault,
	author    = {Koch, Simon and Klein, David and Johns, Martin},
	title     = {The Fault in Our Stars: An Analysis of GitHub Stars as an Importance Metric for Web Source Code},
	booktitle = {Workshop on Measurements, Attacks, and Defenses for the Web (MADWeb)},
	year      = {2024},
	pages     = {65--76},
	url       = {https://www.ias.cs.tu-bs.de/publications/GithubTranco.pdf}
}

@inproceedings{xiao2023early,
	author    = {Xiao, Wenxin and He, Hao and Xu, Weiwei and Zhang, Yuxia and Zhou, Minghui},
	title     = {How Early Participation Determines Long-Term Sustained Activity in GitHub Projects?},
	booktitle = {Proceedings of the 31st ACM Joint European Software Engineering Conference and Symposium on the Foundations of Software Engineering},
	year      = {2023},
	pages     = {29--41},
	publisher = {Association for Computing Machinery},
	address   = {New York, NY, USA},
	doi       = {10.1145/3611643.3616349},
	url       = {https://doi.org/10.1145/3611643.3616349}
}

@inproceedings{ray2014large,
	author    = {Ray, Baishakhi and Posnett, Daryl and Filkov, Vladimir and Devanbu, Premkumar},
	title     = {A Large Scale Study of Programming Languages and Code Quality in GitHub},
	booktitle = {Proceedings of the 22nd ACM SIGSOFT International Symposium on Foundations of Software Engineering},
	year      = {2014},
	pages     = {155--165},
	publisher = {Association for Computing Machinery},
	address   = {New York, NY, USA},
	doi       = {10.1145/2635868.2635922},
	url       = {https://doi.org/10.1145/2635868.2635922}
}

@article{kaushik2026,
	author    = {Kaushik, Mohit and Chahal, Kuljit Kaur},
	title     = {Community Engagement and the Lifespan of Open-Source Software Projects},
	journal   = {Information and Software Technology},
	volume    = {189},
	pages     = {107914},
	year      = {2026},
	issn      = {0950-5849},
	url       = {10.1016/j.infsof.2025.107914}
}

@inproceedings{dey2020exploratory,
	author    = {Dey, Tapajit and Vasilescu, Bogdan and Mockus, Audris},
	title     = {An Exploratory Study of Bot Commits},
	booktitle = {Proceedings of the IEEE/ACM 42nd International Conference on Software Engineering Workshops},
	year      = {2020},
	pages     = {61--65},
	publisher = {Association for Computing Machinery},
	address   = {New York, NY, USA},
	doi       = {10.1145/3387940.3391502},
	url       = {https://doi.org/10.1145/3387940.3391502}
}

\end{document}